\documentclass[longauth]{aa}  

\usepackage{graphicx}
\usepackage{txfonts}
\usepackage[authoryear]{natbib}
\begin{document}
   \title{Discovery of the two ``wings'' of the Kookaburra complex in VHE
$\gamma$-rays with H.E.S.S.}

\author{F. Aharonian\inst{1}
 \and A.G.~Akhperjanian \inst{2}
 \and A.R.~Bazer-Bachi \inst{3}
 \and M.~Beilicke \inst{4}
 \and W.~Benbow \inst{1}
 \and D.~Berge \inst{1}
 \and K.~Bernl\"ohr \inst{1,5}
 \and C.~Boisson \inst{6}
 \and O.~Bolz \inst{1}
 \and V.~Borrel \inst{3}
% ...
}
\institute{
Max-Planck-Institut f\"ur Kernphysik, P.O. Box 103980, D 69029
Heidelberg, Germany
\and
 Yerevan Physics Institute, 2 Alikhanian Brothers St., 375036 Yerevan,
Armenia
\and
Centre d'Etude Spatiale des Rayonnements, CNRS/UPS, 9 av. du Colonel Roche, BP
4346, F-31029 Toulouse Cedex 4, France
\and
Universit\"at Hamburg, Institut f\"ur Experimentalphysik, Luruper Chaussee
149, D 22761 Hamburg, Germany
\and
Institut f\"ur Physik, Humboldt-Universit\"at zu Berlin, Newtonstr. 15,
D 12489 Berlin, Germany
\and
LUTH, UMR 8102 du CNRS, Observatoire de Paris, Section de Meudon, F-92195 Meudon Cedex,
France
% ...
}





   \date{}

 
  \abstract 
  {} 
  {Search for  Very High Energy $\gamma$-ray  emission in the
  Kookaburra complex through  observations with the H.E.S.S. array.}  
  {Stereoscopic imaging of Cherenkov
  light emission of the $\gamma$-ray showers in the atmosphere is used
  for the reconstruction and selection of the events to
  search for  $\gamma$-ray signals. Their spectrum is derived by a
  forward-folding maximum likelihood fit.}
  {Two extended $\gamma$-ray sources with an angular (68\%) radius of $3.3-3.4${\arcmin}  are
  discovered at high ($>$13$\sigma$) statistical
  significance: HESS~J1420-607 and HESS~J1418-609. They exhibit a flux above 1 TeV of ($2.97 \pm 0.18_{\rm
stat}  \pm 0.60_{\rm sys})
  \times 10^{-12}$ and ($2.17 \pm 0.17_{\rm stat} \pm 0.43_{\rm sys}) \times 10^{-12}$
  cm$^{-2}$~s$^{-1}$, respectively, and similar hard photon indices $\sim 2.2$. 
Multi-wavelength comparisons show spatial coincidence with the
wings of the Kookaburra.  Two pulsar wind nebul{\ae} candidates,  
K3/PSR J1420-6048 and the Rabbit, lie on the edge of the H.E.S.S. sources.}
  { The two new sources confirm  the non-thermal nature
of at least parts of the two radio wings which overlap with the $\gamma$-ray emission 
and establish their connection with the two X-ray pulsar wind nebul{\ae} candidates.
Given the large point spread function of EGRET,
the unidentified source(s)
3EG~J1420$-$6038/GeV~J1417$-$6100 could possibly be related to either or both H.E.S.S. sources.
The most likely explanation for the 
Very High Energy $\gamma$-rays discovered by H.E.S.S. is inverse Compton emission
of accelerated electrons on the Cosmic Microwave Background near the two candidate 
pulsar wind nebul{\ae}, K3/PSR J1420-6048 and the
Rabbit.  Two scenarios which could
lead to the observed large ($\sim$10 pc) offset-nebula type morphologies
are briefly discussed.
}

    \titlerunning{Discovery of two sources in the Kookaburra region with H.E.S.S.}

   \maketitle
%

\section{Introduction}

\begin{figure*}[t]
  \centering
  \includegraphics[width=0.5\textwidth]{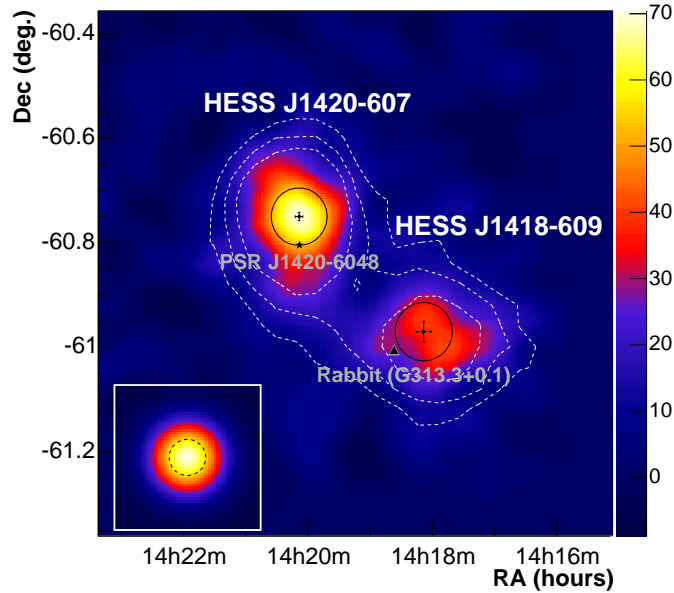}
  \caption{Smoothed excess map of the $1^{\circ} \times 1^{\circ}$
  field of view in the Kookaburra region. The unbinned excess map has
  been smoothed with a Gaussian of width 2{\arcmin}. In the bottom left
  corner the point-spread function of this dataset smoothed in the
  same way is shown (along with the smoothing radius as a black dashed
  line). The white contours denote the 5$\sigma$, 7.5$\sigma$ and
  10$\sigma$ significance levels (with the outermost being the
  5$\sigma$ contour), derived with a point source integration radius
  of $0.1^{\circ}$. The position of the pulsar PSR\,J1420$-$6048 is
  marked with a star, the position of the rabbit (G313.3+0.1) is
  marked with a black and white triangle. The best fit positions of the
  two sources are marked with error crosses, the best-fit extensions
  are given as black circles.There is no significant evidence for a connecting 
  bridge beyond what is expected from the Gaussian source shape
  convoluted with the point spread function of the
  instrument.}
  \label{fig::skymap}
\end{figure*}

The complex of compact and extended radio/X-ray sources called
Kookaburra, after the name of the Australian bird~\citep{rrjg99},
spans over about one square degree along the Galactic plane around
$l=313.4^{\circ}$. It has been extensively studied in the search for
counterparts to the unidentified EGRET source (or sources, see
section 3.3) 3EG~J1420$-$6038/GeV~J1417$-$6100~\citep{ht99,lm97}. 

Radio images of this region
have revealed a  large circular thermal shell with a broad wing
to the North-East and a narrower wing to the South-West.  
Diffuse X-ray emission in the wings and point sources have been discovered through
ASCA observations~\citep{rrjg99,rrj01}, followed by higher resolution
imaging with XMM-{\it Newton} and {\it Chandra}~\citep{ng05}. 
A 3$^\prime$  X-ray/radio nebula in the North-Eastern wing contains a
young ($\tau_c=1.3 \times 10^4$ yr) and  
very energetic radio pulsar (${\dot E} = 1.0\times
10^{37}$erg/s), PSR J1420$-$6048~\citep{dam01}, and has been proposed 
as a candidate pulsar wind nebula (PWN) counterpart to the EGRET
source(s).   
A brighter 
nebula, G313.1+0.1, called the `Rabbit' 
\citep{rrjg99}, lying in the South-Western wing and exhibiting 
extended
hard X-ray
emission has been proposed as a plausible PWN contributing also to the 
$\gamma$-ray emission detected by EGRET.

In this paper, observations of the Kookaburra region with the H.E.S.S. (High Energy Stereoscopic System) 
telescopes and 
the discovery of two Very High Energy (VHE) $\gamma$-ray sources
coinciding with its wings are reported. 
H.E.S.S. is an array of four
imaging atmospheric Cherenkov telescopes located in the Khomas
Highland of Namibia~\citep{HESS}. 
Each H.E.S.S. telescope has a mirror area of 107 m$^2$
~\citep{HESSOptics} and a total field of view of 5$^\circ$\citep{HESSCamera}.
The system is run in a coincidence
mode~\citep{HESSTrigger} requiring at least two of the four telescopes
to have triggered in each event. The H.E.S.S. instrument has an
angular resolution of 
$\sim 5^{\prime}$
per event and a point-source
sensitivity of $<2.0\times\,10^{-13}$ cm$^{-2}$s$^{-1}$ (1\% of the
flux from the Crab Nebula above 1 TeV) for a $5\,\sigma$ detection in a 25 hour
observation.

Section 2 describes the  H.E.S.S. observations, data reduction and
results. In section 3 multi-wavelength comparisons are made and in the
following section  the
interpretation of data together with possible associations for the new
H.E.S.S. sources are discussed. 

\begin{figure*}
  \centering
  \includegraphics[width=0.75\textwidth]{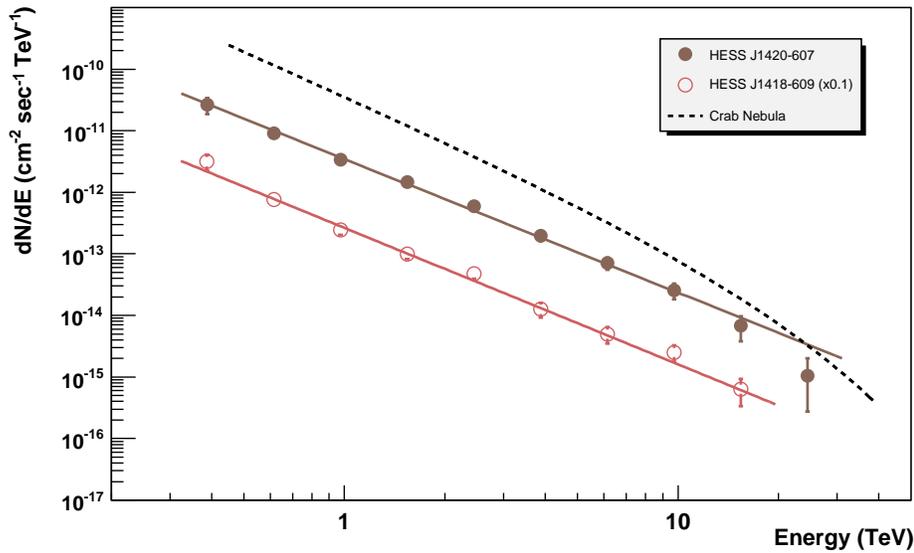}
  \caption{Reconstructed differential VHE $\gamma$-ray spectrum of
  HESS\,J1420$-$607 (full circles) and HESS\,J1418$-$609 (open circles)
  along with power law fits to the two spectra. The flux points for
  HESS\,J1418$-$609 have been scaled by a factor of 0.1 for ease of
  viewing. The fit to the HESS\,J1420$-$607 data results in a photon
  index of $2.17\pm0.06_{\rm stat}\pm 0.1_{\rm sys} $ with a flux normalisation at 1 TeV of
  $(3.48 \pm 0.20_{\rm stat}\pm 0.70_{\rm sys}) \times 10^{-12} \mathrm{cm}^{-2} \,
  \mathrm{s}^{-1} \, \mathrm{TeV}^{-1}$. The power-law fit for
  HESS\,J1418$-$609 yields a photon index of $2.22 \pm 0.08_{\rm stat}\pm 0.1_{\rm sys}$ and
  a flux normalisation of $(2.64 \pm 0.20_{\rm stat} \pm 0.53_{\rm sys})\times 10^{-12}
  \mathrm{cm}^{-2} \, \mathrm{s}^{-1} \, \mathrm{TeV}^{-1}$. The dashed line shows the 
  Crab nebula spectrum as measured by H.E.S.S.~\citep{HESSCrab}.}
  \label{fig::spectrum}
\end{figure*}


\section{H.E.S.S. Observations}
The first observations of the Kookaburra region took place in a survey
of the Galactic plane in the range of Galactic longitudes $300^{\circ}
< l < 330^{\circ}$ and Galactic latitudes $-3^{\circ} < b <
3^{\circ}$. This survey, which 
was taken
between April and July 2005,
represents the extension of the H.E.S.S. 2004 survey of the inner
Galaxy~\citep{HESSScan,HESSScanII} toward lower Galactic longitudes.
The detection of a $\gamma$-ray signal from the Kookaburra region
triggered re-observations between May and August 2005 for 11.1 hours
in pointed observations alternating at an offset of 0.7$^\circ$ in
declination around a central position in the Kookaburra (14h20m0s,
-60d45\arcmin). The average zenith angle of observations was
35.3$^{\circ}$. The dead-time corrected 
data set
amounts to 18.1 hours within $2^{\circ}$ of the central position in
the Kookaburra.

After calibration, the standard H.E.S.S. event
reconstruction scheme was applied to the data (see ~\citet{HESS2155}
for details).  Cuts on the scaled width and length of images
(optimised on $\gamma$-ray simulations and off-source data) are used to
suppress the hadronic background. As previously
described~\citep{
HESSScanII}, two different sets of image size cuts are
applied: to study the morphology of a source and achieve a maximum
signal-to-noise ratio for a weak source with a hard spectrum, a rather
tight 
cut on the number of photoelectrons (p.e.) in the image of 200 p.e.
is applied. For the spectral analysis
this image size cut is loosened to 80~p.e. to achieve a maximum
coverage in energy, resulting in a spectral analysis threshold of
300~GeV for the dataset described here. 
Different methods for deriving a
background estimate as described in~\citet{HESSBack} are applied. For the spectral
analysis, the background is usually taken from positions in the field
of view with the same offset from the pointing direction as the source
region to 
obliviate the
need for corrections concerning the radial dependence of the
background acceptance.
The background estimate for each position in the
two-dimensional sky map is 
taken from a ring of mean radius
0.7$^{\circ}$ and an area  
seven times that of the on-source region. In all
background estimation methods, known $\gamma$-ray emitting regions are
excluded from the background regions to avoid $\gamma$-ray
contamination of the background estimate (after iterations for newly discovered sources).
It should be noted that
consistent results are achieved with different background estimation
techniques. Figure~\ref{fig::skymap} shows a smoothed excess counts map
of the Kookaburra region along with contours that correspond to
5$\sigma$, 7.5$\sigma$, and 10$\sigma$ significance levels.

Two sources of very high energy $\gamma$-rays are apparent in this
map at high statistical significance. The stronger of the
two sources,
HESS\,J1420$-$607,
extends to the North of the energetic pulsar
PSR\,J1420$-$6048. 
Slices in different directions
through the source show a symmetric profile with consistent
extensions. Therefore, the assumption of a radially symmetric Gaussian
emission region ($\rho \propto \exp(-\theta^{2}/2\sigma^2)$ with
$\sigma^2=\sigma_{\mathrm{PSF}}^2+\sigma_{\mathrm{source}}^2$,
$\sigma_{\mathrm{PSF}}$ characterizing the point spread function) 
seems well
justified.
 With this assumption an intrinsic extension of
$\sigma_{\mathrm{source}}\,=\,3.3$\arcmin $\pm$0.5{\arcmin} is derived.
The best-fit position for the centre of the excess lies at 
14h20m9s$\pm$4s,
$-$60d45\arcmin 36\arcsec$\pm$32\arcsec. The slightly less bright second source,
HESS\,J1418$-$609,
at a distance of $\sim 33{\arcmin}$ to the South-West from
HESS\,J1420$-$607 extends to the West of the Rabbit,
at 14h18m4s$\pm$7s, $-$60d58\arcmin 31\arcsec $\pm$ 35{\arcsec}
at a distance of 8.2{\arcmin} to the position of the Rabbit. The best
fit extension of HESS\,J1418$-$609 is
$\sigma_{\mathrm{source}}\,=\,3.4$\arcmin $\pm$0.6\arcmin. Fitting an
elongated Gaussian shape to this source yields a semi-major axis of
$4.9\arcmin\pm 1.5{\arcmin}$ and a semi-minor axis of $2.7\arcmin \pm
0.7{\arcmin}$ at 
a position angle
of $46.2^{\circ} \pm 20.4^{\circ}$ (major axis, North to East). 
Using the positions derived
above and applying a cut on the reconstructed angular distance of
$\gamma$-ray candidate events from this best-fit position of $\theta <
0.16^{\circ}$, yields a statistical (pre-trial) significance of 15.2$\sigma$ at
an excess of $692 \pm 26$ events for HESS\,J1420$-$607 and a statistical
(pre-trial) significance of 13.2$\sigma$ and an excess of 576$\pm$24 events for
HESS\,J1418$-$609. The post-trial significances differ by less than 0.5$\sigma$.
There is no significant evidence for a connecting bridge beyond
what is expected from the Gaussian source shape
convoluted with the point spread function of the
instrument. Using the fit function, the
contamination of HESS\,J1420$-$607 in the integration circle of
HESS\,J1418$-$609 was estimated to be 3\%.

\begin{figure*}
  \centering
  \includegraphics[width=0.49\textwidth]{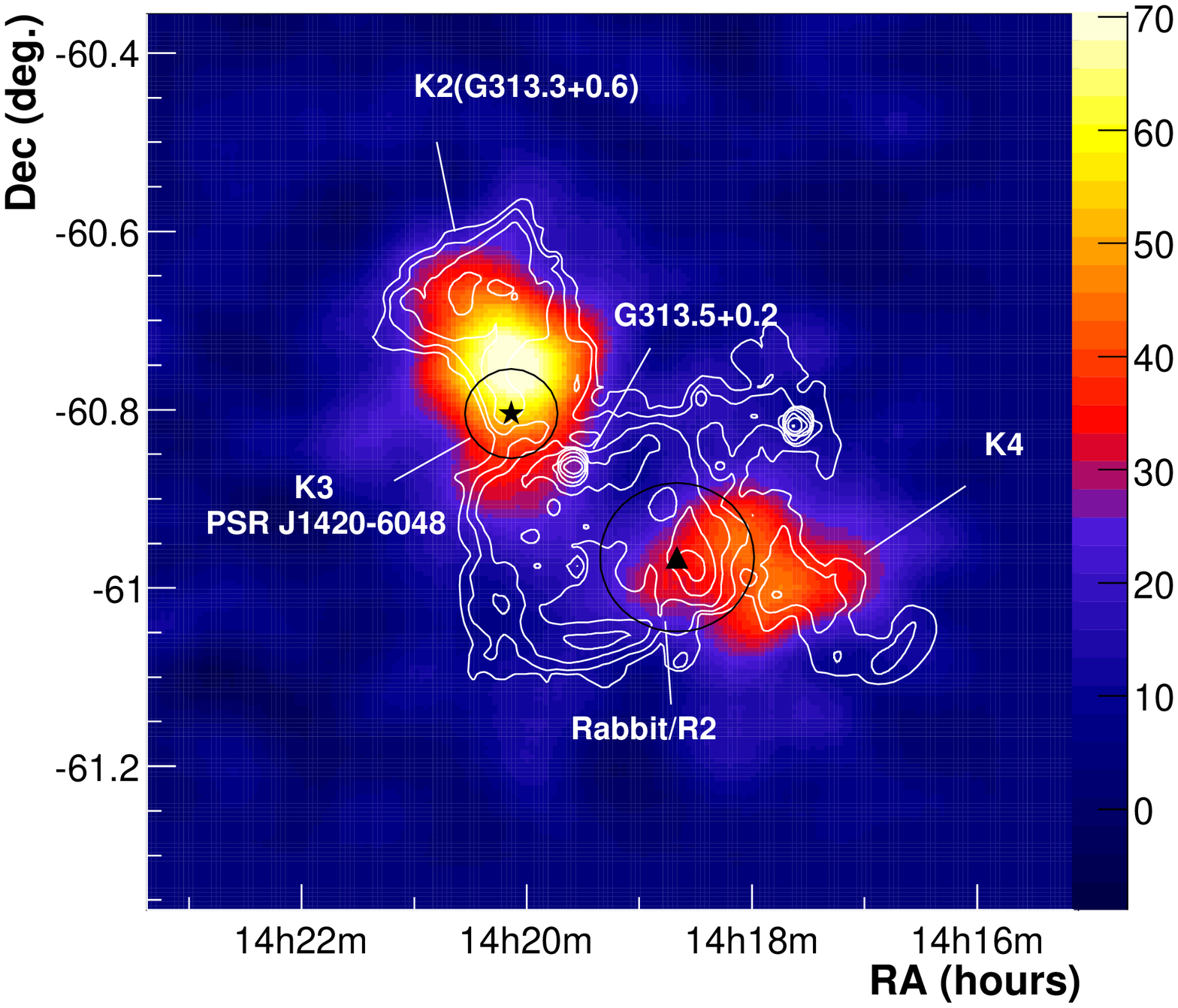}
  \includegraphics[width=0.49\textwidth]{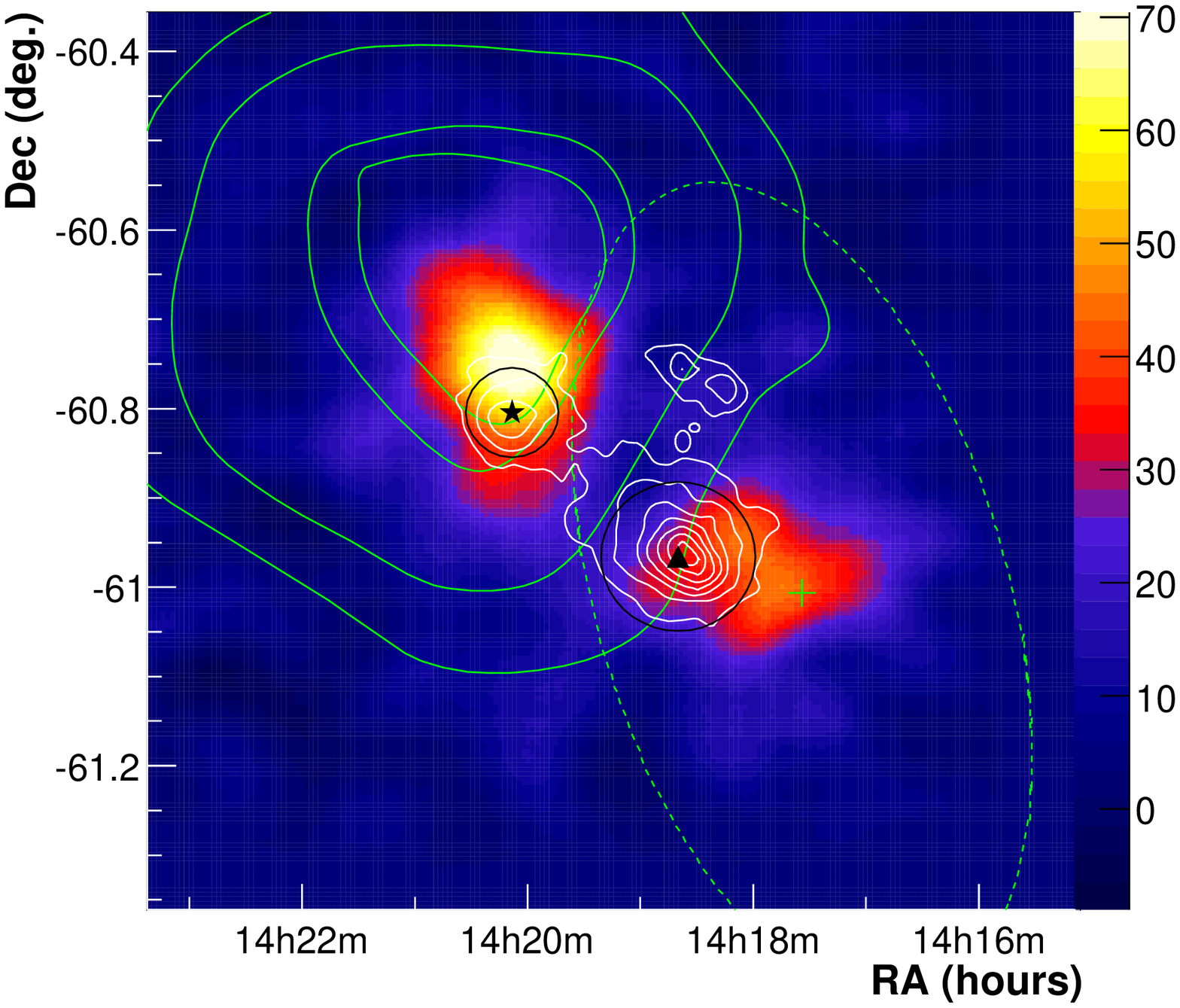}
  \caption{H.E.S.S. excess map smoothed with a Gaussian of $\sigma=2^\prime$. 
	The black star and triangle mark the positions of the pulsar 
  	PSR\,J1420$-$6048 and the Rabbit (G313.3+0.1) R2 source (see text). Black
	circles around these two positions show the approximate extension of the 
	X-ray diffuse emission for K3 ($3^{\prime}$) and the Rabbit ($5^{\prime}$) nebula~\citep{ng05}.
	{\it Left panel:} White
	contours are from ATCA 20 cm high resolution images.  
	The radio wings, K2 and K4
	are clearly correlated with the H.E.S.S. map, whereas there is
	no correspondence between the central shell, or the bright HII region, 
	G313.5+0.2, and the VHE $\gamma$-ray emission. 
	{\it Right panel:} ASCA GIS
	high energy band data are shown as white contours. 
	Green contours show $>$1 GeV confidence levels (50, 68,
	95 and 99\%) for 3EG~J1420$-$6038;
	the green dashed ellipse 
	and the green cross mark the 95\% error box and the postion 
	of GeV~J1417$-$6100. Note that although both EGRET sources are shown, 
	they can not be considered as independent sources (see text). 
	 }
  \label{fig::mwlA}
\end{figure*}

The flux of HESS\,J1420$-$607 and of HESS\,J1418$-$609 have been determined 
both within a
radius of $0.16^{\circ}$, to avoid any overlap in the integration
regions for the two sources. 
 The effective areas used in the
determination of the energy spectra assume full containment of the
source in the integration region.The background has been extracted
from regions distributed on a ring with the same radius and the same offset from the
pointing direction as the integration region.   
The energy estimation algorithm takes into account the optical efficiency change with time 
(characteristic timescale of years as monitored by muon images) as described in~\citet{HESSCrab}. 
The energy spectra of
the two sources have been derived using a forward-folding maximum likelihood
fit~\citep{CATSpectrum} and are very similar, as seen in
Fig.~\ref{fig::spectrum}. The energy spectrum of HESS\,J1420$-$607 can
be fitted with a power-law with a photon index of $2.17 \pm
0.06_{\rm stat}\pm 0.1_{\rm sys}$ and a flux above 1~TeV of ($2.97 \pm
0.18_{\rm stat} \pm 0.60_{\rm sys})\times 
10^{-12}$ cm$^{-2}$~s$^{-1}$ (corresponding to 13.0\% of the flux from
the Crab nebula above that energy). The photon index for
HESS\,J1418$-$609 has a 
similar value of $2.22 \pm 0.08_{\rm stat}\pm 0.1_{\rm sys}$,
the flux of this source above 1~TeV has a value of $(2.17 \pm
0.17_{\rm stat} \pm 0.43_{\rm sys}) \times 10^{-12}$ cm$^{-2}$~s$^{-1}$ (corresponding to
9.6\% of the Crab flux above that energy).

\section{Multi-wavelength search for counterparts}

To search for counterparts, published multi-wavelength data have been
selected and are overlaid on H.E.S.S. excess
maps in the two panels of Fig.~\ref{fig::mwlA}.   
The left panel shows radio contours from 
Australia Telescope Compact Array (ATCA, 20 cm high resolution image) 
and labels of relevant radio sources~\citep{rrjg99}.
Black circles show the approximate extension of the two candidate PWNe
in Kookaburra, K3/PSR~J1420$-$6048 ($3^{\prime}$) and the Rabbit ($5^{\prime}$) in X-rays~\citep{ng05}.  
In the right panel, both 3EG~J1420$-$6038 and GeV~J1417$-$6100, are
shown (confidence contours and error box, respectively) although they are
not independent sources (see section 3.3). X-ray data contours are
taken from ASCA/GIS~\citep{rrk01}.

\subsection{The NE wing and K3/PSR~J1420$-$6048}

HESS~J1420$-$607 is in positional coincidence with the North-Eastern radio wing 
of Kookaburra, G313.3$+$0.6, or K2, which has a rectangular $\sim 12^{\prime} 
\times  8^{\prime}$  shape (at a minimum emission level of 4 mJy/beam).
K2 has a total flux density at 20~cm $S_{20\rm cm}\sim 1$ Jy,
with  
a slight enhancement, labeled K3, of$\sim20$ mJy around  
 PSR J1420$-$6048~\citep{rrjg99}. This young  68.2~ms pulsar~\citep{dam01} which shows a high  
spin-down luminosity of ${\dot E} = 1.0\times 10^{37}$erg/s, lies to the South of HESS~J1420$-$607 at 
an angular distance of $\sim 3.1^\prime$. Given the dispersion measure for the pulsar, the 
\citet{cl02} Galactic electron density model implies a distance
$d=5.6\pm0.8$~kpc, closer than its initial estimated distance,  $d\approx 7.7\pm
1.1$~kpc based on \citet{tc93}.
Observations with ASCA and {\it Chandra} have revealed a rather hard nebular X-ray emission 
around PSR J1420$-$6048 extending to a radius $\sim 6^{\prime}$, 
e.g., ~\citet{rrk01} report a 2$-$10~keV ASCA flux of $4.8 \times 10^{-12}
{\rm erg/cm^2s^{-1}}$ with a power-law index $\Gamma =1.4\pm 0.4$ for a  
fitted column density $N_H \sim 1.8 \times 10^{22} {\rm cm^{-2}} $.

The radio spectral indices of K3 and K2 are poorly constrained
(20 to 36 cm: $\alpha_{20/36}=-0.4\pm0.5$ and $\alpha_{20/36}=-0.2\pm0.2$,
respectively~\citep{rrjg99}), but are  
rather hard and compatible with typical PWNe ($-0.3\la \alpha_{20/36} \la
-0.1$). 
The non-thermal nature of at least parts of 
K2 is supported
by the lack of correlation with infrared emission, 
as well as by the X-ray hard diffuse features
and now established by the TeV emission detected by H.E.S.S.
\citet{ng05} have proposed K3 as a candidate PWN and a possible
counterpart to GeV 1417$-$6100, although the nebula detected by {\it
Chandra} in the inner part of K3 is unexpectedly faint.

\subsection{The SW radio wing and the Rabbit nebula}
HESS J1418$-$609 coincides spatially with the narrower $\sim 15^{\prime} 
\times  4^{\prime}$ radio
wing to the South-West, labelled K4, while the Rabbit nebula, or G313.3+0.1,
lies on its Eastern edge at a distance of $8.2^\prime$ to the fitted TeV
position. 
K4 and the Rabbit nebula have each a radio flux of
$S_{20\rm cm}\sim 400$ mJy, with  non-thermal spectral indices (13 to 36 cm)
of $-1.2 \pm 0.5$ and $-0.25 \pm 0.1$, respectively~\citep{rrjg99}. 

In X-rays, the Rabbit nebula is brighter than K3, exhibiting for example in ASCA a
2$-$10~keV  flux of $7.33 \times 10^{-12}{\rm erg/cm^2s^{-1}}$~\citep{rrk01},  
and a power-law index in the range 1.5--1.9.   
Two sources, labelled R1 and R2 (spaced by $42.5 $\arcsec), were resolved by 
{\it Chandra} observations within the diffuse emission 
which extends over an area of radius $\sim 
7^\prime$ ~\citep{ng05}. 
These authors report a very marginal detection (chance probability of
0.02) of X-ray
pulsations from the fainter source, R2, at a period of 108~ms and
propose it as a plausible pulsar for the candidate Rabbit PWN. 
Tentative estimates of the pulsar spin-down luminosity, ${\rm Log}({\dot
E})\sim36.7$ to 37.5 ${\rm erg/s}$, age $\tau=1.6$ kyr  and distance,
$d \sim 5$ kpc are also proposed in \citet{ng05}.

\begin{figure*}
  \centering
  \includegraphics[width=0.75\textwidth]{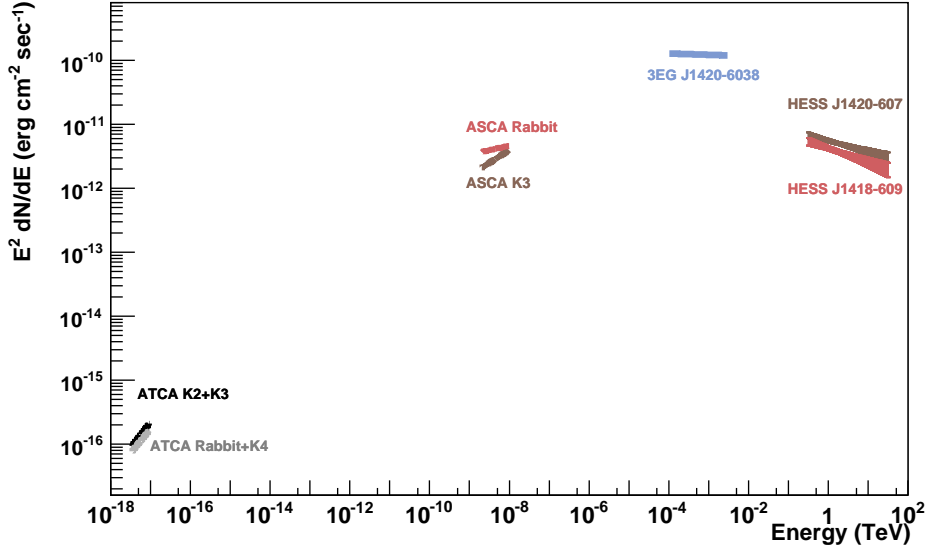}
 \caption{Spectral energy distribution for
	HESS~J1420$-$607/K3/K2 and  HESS~J1418$-$609/Rabbit/K4 : TeV fitted
	spectra are shown as 
	brown and red error boxes, respectively.
	The 3EG~J1420$-$6038 energy flux is
	plotted in blue. 
	X-ray spectra are given for overall nebul{\ae} following the
	discussion in sections 3.1 and 3.2 (ASCA measurements of
	~\citet{rrk01}), in 
	brown and red for K3 and the
	Rabbit, respectively.
	For the radio continuum, although the wings could only be partly associated 
	to the TeV sources,
	the $S_{20\rm cm} \sim 1$ Jy flux of K2/K3 (black), assuming the K3 spectral index $\alpha_{20/36}=-0.2$, 
	and the
	total Rabbit+K4 flux, $S_{20\rm cm} \sim 0.8$ Jy (gray), are plotted. As the radio spectrum of K4 is
	poorly constrained ($\alpha_{20/36}=-1.2 \pm0.5$) the Rabbit nebula index  
	$\alpha_{20/36}=-0.25$ has been used. 
	Note that different integration radii are used (see text) and
        that the X-ray fluxes may be significantly contaminated by contribution from the compact nebular emissions.	
	 }
  \label{fig::vFv}
\end{figure*}

\subsection{The EGRET data}
The original $\gamma$-ray source detected by EGRET in the Kookaburra region,
2EG~J1412-6211, was reported in the second catalog~\citep{th95}, based on the first and 
second year of CGRO operation. 
During phase 3 a new source, 2EGS~J1418$-$6049, was detected and published in the
supplement to the second EGRET catalog~\citep{th96}.  
With the inclusion of phase 4 data, as well as due to the improved
understanding of instrumental responses and backgrounds,  
these two sources evolved to 3EG~J1410-6147 (not shown in Fig.~\ref{fig::mwlA}) and
 3EG~J1420$-$6038, respectively,
in the third EGRET catalog~\citep{ht99}. The latter source exhibits a hard spectral
index, $\Gamma =2.02\pm 0.14$, and a flux above 100 MeV of
$73.8\pm12.1 \times 10^{-8} {\rm cm^{-2}s^{-1}}$, similar to values
reported for 3EG~J1410$-$6147.  Due to their small nominal distance as
compared to the point spread function (PSF) of the instrument and the
above-mentioned similarity, these sources are confused and were
accordingly graded as ''C''.
An analysis above 1 GeV~\citep{lm97} of the EGRET data from the
first 4 years of CGRO operation, where the
higher energy of the $\gamma$-ray events results in a narrower PSF,
yielded only one source in the Kookaburra region, 
GeV~J1417$-$6100, with a flux ($>$1 GeV) of
$9.8\pm1.8 \times 10^{-8}{\rm cm^{-2}s^{-1}}$.
In Fig.~\ref{fig::mwlA}, both 3EG~J1420$-$6038 (confidence 
contours) and GeV~J1417$-$6100 (error box and cross) are shown. HESS~J1420$-$607
lies at an angular distance of $7.5^\prime$ 
($\theta_{95}^{\rm EGRET}\sim20^{\prime}$) to the former, whereas    
the position reported for GeV~J1417$-$6100 is at $4^\prime$ of HESS J1418$-$609
($\theta_{95}^{\rm EGRET}\sim31^{\prime}$).
However, as even above 1 GeV the containment radius of EGRET PSF
is quite large, GeV~J1417$-$6100 and 3EG~J1420$-$6038 share photons and can not be considered 
as independent sources. 
Another feature of the EGRET data is the indication of
variability reported by~\citet{nol03} for
3EG~J1420$-$6038. Comparing the estimated level of variability 
with the average for EGRET PWN candidates, these authors 
suggest the possible contribution of a PWN component to the $\gamma$-ray emission above 100 MeV 
of 3EG~J1420$-$6038.

\section{Discussion}

\subsection{Morphology: large offset nebul{\ae}}
Following the multi-wavelength discussion of the previous section, the
two discovered VHE sources are most plausibly associated to the two  
candidate PWNe, K3/PSR~J1420$-$6049 and Rabbit/R2. 
The extent of the
TeV sources and their spatial coincidence with the radio wings, K2 and K4, as
well as the non-thermal properties of the latter,
strongly suggest that the wings are at least partly related to the two PWNe. 
This connection and the respective positions of K3 and the Rabbit on
the edges of the two H.E.S.S. sources imply in turn an
asymmetric/offset-nebula type configuration, similar to that of the
rapidly moving PWNe (or RPWNe), e.g. the Mouse nebula G359.23$-$0.3, or
analogous 
to one-sided ``crushed'' nebul{\ae}, e.g. Vela X or G18.0$-$0.7. 
Both of the latter objects have recently been associated
to VHE sources, HESS J0836$-$456~\citep{HESS0836} and
HESS J1825$-$137~\citep{HESS1825}, respectively.  

Given the distance estimate for PSR J1420$-$607, $d=5.6 d_{5.6}
$~kpc, the measured (2$\sigma$) 
angular extension of  HESS~J1420$-$607 yields a relatively large
nebular 
projected diameter
of $11 d_{5.6}$ pc.
For HESS~J1418$-$609 and the Rabbit, the rough estimate of the distance of R2,
$d \sim 5 d_{5}$~kpc, would imply a larger, but still
comparable size of $14 d_{5}$ pc $\times\,8 d_{5}$ pc. 
The implications of such large extensions will be discussed briefly in section 4.3.

\subsection{Spectral Energy Distribution}
Fig.~\ref{fig::vFv} shows tentative broadband spectral energy distributions (SED)
of the H.E.S.S. sources, assuming their association with the radio/X-ray 
nebul{\ae} and with the wings. 
X-ray spectra are plotted for overall nebul{\ae} following the
discussion in sections 3.1 and 3.2 (ASCA measurements of
~\citet{rrk01}), and can be significantly contaminated by the
compact nebula hard emission.
For the radio continuum, although the wings extend further than the TeV sources,
the $S_{20\rm cm} \sim 1$ Jy flux of K2/K3
and the total Rabbit+K4 flux, $S_{20\rm cm} \sim 0.8$ Jy, are plotted. 
Given the discrepancy in spatial resolution between $\gamma$-ray
observations with EGRET and H.E.S.S., we indicate the energy flux
of 3EG~J1420$-$6038 for both H.E.S.S. sources.
Assuming a pulsar-only origin first, the inferred conversion efficiency of spin-down
power to pulsed luminosity in the 100 MeV to 10 GeV (assuming a
typical Vela-like pulsed cutoff of 10 GeV) is $< 1.6\times
10^{-2}f$, with the unknown beaming factor $f=(\Delta \Omega/\beta)\%$
the ratio of the gamma-ray beaming solid angle to the pulse duty cycle
$\beta$, while the ``$<$'' sign refers to the sharing of photons
in the case of two unresolved overlapping EGRET sources. With a typical $f\sim
1$, the inferred conversion efficiency would then be of the same order of
magnitude as that of the Vela (EGRET) pulsar for the same energy
band. The other possibility, favoured by the indication of
variability in EGRET data (see section 3.3), is the contribution of  
a plerionic component
in which case, given 
the similarity of the flux/spectra of the H.E.S.S. sources, the overall
EGRET flux would include contributions from both HESS~J1420$-$607 and
HESS~J1418$-$609.  

In a plerionic scenario, given the increasing order 
of synchrotron loss time-scales for
X-ray, VHE, GeV and radio emitting electrons and the different
integration radii for spectral measurements, the SEDs
could reflect 
the synchrotron/inverse Compton (IC) emission of 
different populations of particles in
different regions. 

For the following discussion, a typical PWN field strength of
$B\sim10^{-5} B_{-5}$~G is assumed and the Cosmic Microwave Background
(CMBR) is chosen as dominant target photons. The synchrotron lifetime
of parent electrons which produce $\gamma$-rays by IC scattering of
these photons, with $E_{e^{-},\rm IC}\approx 20 E_{\rm TeV}^{1/2}$
TeV, is $\tau(E_{e^{-},\rm IC})\approx 4.8 B^{-2}_{-5} E^{-1/2}_{\rm
TeV}$ kyr.
This yields, for the mean $\bar E_{\rm TeV} \sim 0.8$~TeV and  $\bar
E_{\rm GeV} \sim 0.5$ GeV energies,  $\tau(E_{e^{-}})\sim$ 5 and
$\sim$ 200 kyr, respectively. 
Thus, whereas the electron lifetimes of VHE emitting electrons can be
comparable to the ages of these two PWNe, the radio and GeV emitting
lifetimes should be much longer. Only the X-ray emitting lifetimes
should be much shorter: 
$\tau(E_{e^{-},\rm
Syn})=1.2 B^{-3/2}_{-5} E^{-1/2}_{\rm keV}$ kyr, where $E_{e^{-},\rm
Syn}=70  B^{-1/2}_{-5} E_{\rm keV}^{1/2}$ TeV. 
The hard X-ray emitting electrons in the extended nebula would then
correspond to freshly injected electrons by the pulsar wind shock.
Upon advection away from the pulsar wind shock, the X-ray photon index
should steepen towards the outer nebula as a result of
cooling. However, the measured radio and possibly GeV emissions
are expected to contain
uncooled particles from the earliest stages of the pulsar injection. In
contrast, those emitting at TeV energies (for pulsar ages
$\tau_c \ga 10$ kyr) should be composed mainly 
of cooled particles cumulated for up to 5$-$10 kyr. 
Hence, radio and GeV nebul{\ae} should have larger
sizes than that of the TeV nebula for evolved PWNe 
and the latter should in turn be larger than the hard X-ray diffuse emission.
In this simple picture the difference between the VHE $\gamma$-ray
and the X-ray spectral indices should be $\sim 0.5$.
The measured $\Gamma_{\rm TeV} \sim 2.2$
(determined within a 
radius $\sim10^\prime$ for both nebul{\ae}), when compared to the X-ray
measurements on smaller radii (within few arc-minutes), $\Gamma_{X} \sim 1.5-1.8$, 
are roughly consistent with this picture. One expects also a harder
VHE spectrum relative to the total observed, if events are selected
within the PSF around these two source origins. Such details will
however be addressed in a future paper.
 
Realistically, the situation becomes more complicated when  
magnetic field variations in space/time, evolution of the pulsar spin-down luminosity, or other potential target
photons, 
e.g. a cold dust component, are taken into account. A 
higher field strength in the early epochs would shorten the cooling
timescales, whereas the IC electrons corresponding to dust IR
targets have longer synchrotron lifetimes ($\sim$ factor of two for 25
K targets).

In Table~\ref{table:2}, the $\gamma$-ray  luminosities
of the two sources are compared to the recently discovered TeV source
HESS~J1825$-$137. The
latter, associated with G18.0$-$0.7 and the Vela-like pulsar
PSR~B1823$-$13, shows a similar offset morphology with an even larger
TeV nebula (D$\sim$ 34 pc).
As mentioned above, the TeV luminosity
should reflect the emission of an accumulated population of particles
injected into the nebula through the lifetime of the pulsar.
For similar magnetic field configurations, the apparent $\gamma$-ray efficiency
should be an increasing function of the 
nebula age for young and middle-aged PWN, and may reach large values when derived
with respect to the present day spin-down luminosity. 
The values derived here follow this trend although they suffer from large
uncertainties on distance and age measurements, especially for the
Rabbit. For HESS~J1420$-$609 and/or HESS~J1418$-$607, they may have been
significantly underestimated if the  $\gamma$-ray emission continues
into the EGRET domain.

\begin{table}[htb]
\caption{$\gamma$-ray luminosities, 
efficiencies and ages for the two PWNe as compared to G18.0$-$0.7/HESS~J1825$-$137.
Distances of 5.6, 5 and 4~kpc have been used for  HESS~J1420$-$609,
HESS~J1418$-$607, and HESS~J1825$-$137 respectively.
. } 
\label{table:2}      
\centering                          
\begin{tabular}{l c c c c}        
\hline\hline                 
Source & ${\dot E}$ & $ L_{\gamma \rm}$ & $\epsilon_{\gamma}$ &age \\    
       & $10^{36}$erg/s & $10^{33} {\rm erg/s}$ & \%& kyr\\
\hline  
   K3/ & & & &\\                      
   HESS~J1420$-$609 &  $10$ & 51 & $0.51$& 13\\ 
\hline
   Rabbit/ & & & &\\
   HESS~J1418$-$607 &
$ \sim5-30 $  & 48 & $\sim 0.96-0.16$ &$1.6$?\\
\hline
   G18.0$-$0.7/ & & & &\\
   HESS~J1825$-$137 & $2.8$ & 100& $3.6 $& 21\\
\hline                                   
\end{tabular}
\end{table}

\subsection{Extended one-sided nebul{\ae}  scenarios}

In the following discussion we adopt  $D \sim$ 10 pc for the spatial
extent (projected diameter) of the two H.E.S.S. sources. 
The large sizes of the TeV sources imply very high speeds for
transporting the particles to the edges of the PWNe within their
synchrotron lifetimes,
e.g. $D/ \tau(E_{e^{-},\rm IC}) \sim 2000$ km s$^{-1}$. 
We examine briefly two
scenarios where such large and asymmetric extents can be
expected.

The Mouse nebula, G359.23$-$0.82, is a well studied example of RPWNe,
i.e. an energetic pulsar interacting with its surrounding material:
the supersonic pulsar's velocity confines the nebula through ram 
pressure resulting in a bow-shock structure \citep{mouse04} and an
elongated nebula with the pulsar at its apex. The ``tail'' of
the mouse contains shocked pulsar material convected away at
very high flow velocities, typically fractions of light speed as shown by
relativistic MHD simulations of \citet{bad05}.   
While X-ray exposures
seem too short to reveal fine details for
either PWN candidate~\citep{ng05}, the high resolution ATCA 
radio images of K2/K3
and the Rabbit~\citep{rrj01,rrjg99} show filamentary enhancements which could trace 
North-South or West-East motions, respectively. However, the fact that  
a bright part of the X-ray nebula is to the South of
PSR~J1420$-$6049 while the VHE emission lies mainly to its North argues against  
this interpretation for K3/HESS~J1420$-$607. In the case of the
Rabbit/HESS~J1418$-$609, the uncollimated diffuse X-ray and VHE
emissions may not be consistent with the RPWN scenario, either.  

One-sided PWNe can also be produced in evolved systems in which the reverse
shock from the surrounding SNR displaces the nebula
\citep{rc84}.
During the crushing phase, the nebula particles can be convected  with speeds of  $\sim$1000 $\rm
km s^{-1}$ 
 ~\citep{swa01}
away from the pulsar and feed an offset nebula, if initial asymmetries
in the system (offset of 
the pulsar with respect to the expanding ejecta, density gradients
around the birth site) yield a composite reverse shock with
different arrival timescales to the PWN \citep{ch98}.  
Simulations of  ~\citep{bcf01}  
show that for a symmetrical system and reasonable assumptions 
the start of inward motion of the reverse shock
occurs at  $\sim$ 1500 yr and the crushing takes place on timescales of few
thousand years, 
after which the expected ratio of the pulsar nebula radius to
that of the SNR $R_{\rm PWN}/R_{\rm SNR}$  $\sim$0.25.
The size of the undetected parent SNR for PSR~J1420$-$6048, assuming an age
$\tau \sim$13 kyr and expansion in an environment of density $\sim 
\rm few\, 10^{-1} cm^{-3}$, $R_{\rm SNR}\sim 20-25$ pc  
 as compared to the measured $R_{\rm PWN}$, would fit with the
aforementioned simulations for
HESS~J1420$-$607.
The offset morphology of Rabbit/HESS~J1418$-$609 could also be explained
through this scenario, provided that they constitute an evolved PWN,
i.e. the age of the system is at least few thousand years. In this case
the parent SNR again remains undetected.

\section{Conclusions}

Two extended VHE sources have been discovered in the wings of the Kookaburra complex, HESS~J1420$-$609
and HESS~J1418$-$607. 
They show similar $\gamma$-ray angular extensions ($\sim$
3.3${\arcmin}$-$3.4${\arcmin}) and hard VHE $\gamma$-ray spectra. 
This discovery confirms the
non-thermal nature of at least parts of the wings of the Kookaburra
which overlap with the VHE emission 
and establishes their connection with the two X-ray PWN candidates,
K3  and the Rabbit.  
Within the limits of available multi-wavelength data, the SEDs of the two  
sources show also remarkable similarities and suggest analogous
underlying objects and emission processes. 
The VHE $\gamma$-ray emission could most plausibly be explained by
IC emission from these PWNe in an offset-type configuration
near the energetic PSR~J1420$-$6048, and the candidate pulsar in the
Rabbit nebula.
Given the poor spatial resolution of EGRET measurements,
the unidentified source(s) 3EG~J1420$-$6038/GeV~J1417$-$6100
could possibly be related to either or both H.E.S.S. sources through a PWN-type emission above 100 MeV.
The EGRET flux may also contain pulsed emission from the
pulsar associated with K3 and the candidate pulsar in the Rabbit.
Future GLAST
observations of these objects should be able to establish the
status of such pulsed emission, resulting in a more certain
multiwavelength interpretation of the H.E.S.S. sources.
The detection of the Rabbit pulsar, the morphology of the 
VHE and X-ray nebul{\ae}, and the confirmation of a GeV-TeV connection  
are important considerations that need to be addressed
through further investigations. 


\begin{acknowledgements}
The support of the Namibian authorities and of the University of Namibia
in facilitating the construction and operation of H.E.S.S. is gratefully
acknowledged, as is the support by the German Ministry for Education and
Research (BMBF), the Max Planck Society, the French Ministry for Research,
the CNRS-IN2P3 and the Astroparticle Interdisciplinary Programme of the
CNRS, the U.K. Particle Physics and Astronomy Research Council (PPARC),
the IPNP of the Charles University, the South African Department of
Science and Technology and National Research Foundation, and by the
University of Namibia. We appreciate the excellent work of the technical
support staff in Berlin, Durham, Hamburg, Heidelberg, Palaiseau, Paris,
Saclay, and in Namibia in the construction and operation of the
equipment. The authors wish to thank the referee, M. Roberts, for
very useful comments on the multi-wavelength discussion.
\end{acknowledgements}

\bibliographystyle{aa}

\end{document}